\documentclass[conference,10pt,letter]{IEEEtran}
\usepackage{graphicx}
\usepackage{amsmath}
\usepackage{amssymb}
\usepackage{definice}
\usepackage{calc}
\usepackage{multirow}
\usepackage{mathrsfs}
\usepackage[english]{babel}
\usepackage{float}
\usepackage{cool}
\IEEEoverridecommandlockouts
\usepackage{times}
\usepackage{color,colortbl}
\usepackage{tikz}
\usepackage{caption}
\usepackage{subcaption}
\usetikzlibrary{calc,arrows,positioning}
\definecolor{TableCol}{rgb}{1,0.75,0}
\definecolor{TableCol2}{rgb}{1,0.9,0}
\definecolor{TableCol3}{rgb}{1,0.5,0}
\usepackage{enumitem}
\usepackage{commath}
\usepackage{tensor}
\usepackage{mathtools}
\usepackage[thref]{ntheorem}
\usepackage{textcomp}
\usepackage{tabstackengine}
\usepackage[makeroom]{cancel}

\usepackage{textcomp}

\usepackage{times}
\usepackage{amssymb}

\usepackage{cite} 
\usepackage{stfloats}  

\newcommand{\pderivv}[2]{\frac{\partial^2 #1}{\partial {#2}^2}}

\newcounter{steps}
	{\end{list}}

\usepackage[scr=boondoxo,scrscaled=1.05]{mathalfa}

\def\bfA{\mathbf A}

\def\bfD{\mathbf D}

\def\bfF{\mathbf F}
\def\bfG{\mathbf G}

\def\bfI{\mathbf I}

\def\bfP{\mathbf P}
\def\bfQ{\mathbf Q}
\def\bfR{\mathbf R}
\def\bfS{\mathbf S}
\def\bfT{\mathbf T}

\def\bfg{\mathbf g}
\def\bfh{\mathbf h}

\def\bfr{\mathbf r}

\def\bfv{\mathbf v}
\def\bfw{\mathbf w}
\def\bfx{\mathbf x}
\def\bfy{\mathbf y}
\def\bfz{\mathbf z}

\def\hbfx{\hat{\bfx}}

\def\bfxi{\boldsymbol{\xi}}

\def\bfdelta{\boldsymbol \delta}

\def\bfLambda{\mathbf \Lambda}

\DeclareMathAlphabet\mathbfcal{OMS}{cmsy}{b}{n}

\def\real{\mathbb{R}}

\def\mean{\mathsf{E}}
\def\var{\mathsf{var}}
\def\cov{\mathsf{cov}}

\DeclareMathAlphabet\mathbfcal{OMS}{cmsy}{b}{n}

\begin{document}
	
	\onecolumn
	\vspace*{5cm}
	This paper has been accepted for publication in 2024 27th International Conference on Information Fusion (FUSION).
	Please cite the paper as:
	J. Matoušek, J. Duník and M. Brandner, "Efficient Spectral Differentiation in Grid-Based Continuous State Estimation," 2024 27th International Conference on Information Fusion (FUSION), Venice, Italy, 2024, pp. 1-8, doi: 10.23919/FUSION59988.2024.10706533..
	\clearpage
	\twocolumn
	
	\title{Efficient Spectral Differentiation in Grid-Based Continuous State Estimation}	

\author{J. Matou\v{s}ek, J. Dun\'{i}k, M. Brandner 
	\thanks{Authors are with Department of Cybernetics, Pilsen, Czech Republic. E-mails: e-mail: \{dunikj,matoujak\}@kky.zcu.cz (J. Dun\'{i}k, J. Matou\v{s}ek), and Department of Mathematics, Faculty of Applied Sciences, University of West Bohemia, brandner@kma.zcu.cz (M. Brandner).}
}
	
	\maketitle
	
	\selectlanguage{english}
	\begin{abstract}
		\noindent This paper deals with the state estimation of stochastic models with continuous dynamics. The aim is to incorporate spectral differentiation methods into the solution to the Fokker-Planck equation in grid-based state estimation routine, while taking into account the specifics of the field, such as probability density function (PDF) features, moving grid, zero boundary conditions, etc. The spectral methods, in general, achieve very fast convergence rate of $\mathcal{O}(c^{N}) (O< c < 1)$ for analytical functions such as the probability density function, where $N$ is the number of grid points. This is significantly better than the standard finite difference method (or midpoint rule used in discrete estimation) typically used in grid-based filter design with convergence rate $\mathcal{O}(\frac{1}{N^2})$.  As consequence, the proposed spectral method based filter provides better state estimation accuracy with lower number of grid points, and thus, with lower computational complexity.
	\end{abstract}


	\section{Introduction}
	
	The goal of this paper is to introduce the usage of spectral methods to state estimation, and to show how the spectral methods can benefit from the efficient grid-based point mass filter formulation proposed in \cite{MaDuBrCh:23, DuMaSt:23, MaDuBr:23}.

The continuous state estimation starts with a model 	described by the stochastic differential equation (SDE), and stochastic algebraic equation
\begin{align}
d\bfx(t)&=\bfA \bfx(t)dt+\bfQ d\bfw(t)\label{eq:dynam}\\
\bfz_{t_k}&=\bfh(\bfx_{t_k})+\bfv_{t_k},\label{eq:meas}
\end{align}
where $t$ denotes the time, the vector $\bfx(t)\in\real^{n_x}$ represents the \textit{unknown} state of the system, $\bfz_{t_k}\in\real^{n_z}$ the \textit{known} measurement, and $t_k=t_0,t_1,\ldots,T$ is a discrete time instant, when the measurement arrived. The matrix $\bfA\in\real\\
^{n_x\times n_x}$ and the function $\bfh(\cdot):\real^{n_x}\rightarrow\real^{n_z}$ are known, $\bfQ\in\real^{n_x\times n_x}$ is a known diffusion coefficient, $\bfw(t)$ is the state noise modelled by the Brownian motion with the normally distributed increment with the covariance matrix $E[d\bfw(t)(d\bfw(t))^T]=\bfI_{n_x}dt$, and $\bfv_{t_k}$ is the measurement noise with the known PDF $p(\bfv_{t_k})$. The state and measurement noise random variables are supposed to be independent mutually and of the \textit{known} initial state $\bfx_0$ with the known PDF $p(\bfx_0)$.

The aim of the state estimation (or filtering) is to compute the PDF of the state $\bfx(t)$ conditioned on all past measurements $\bfz^{t_k}=[\bfz_{t_0},\bfz_{t_1},\ldots,\bfz_{t_k}]$. The sought conditional PDF can be either filtering $p(\bfx(t_k)|\bfz^{t_k})$, which is computed in the filter \textit{measurement update} step, or predictive $p(\bfx(t)|\bfz^{t_k}), t>t_k$, which is computed in the filter \textit{time} \textit{update} step.

\subsection{Measurement Update}
The solution to the measurement update providing the filtering PDF is given by the Bayes' rule
\begin{align}
	p(\bfx_{t_k}|\bfz^{t_k})&=\tfrac{p(\bfx_{t_k},\bfz_{t_k}|\bfz^{t_{k-1}})}{p(\bfz_{t_k}|\bfz^{t_{k-1}})}=\tfrac{p(\bfx_{t_k}|\bfz^{t_{k-1}})p(\bfz_{t_k}|\bfx_{t_k})}{p(\bfz_{t_k}|\bfz^{t_{k-1}})},\label{eq:filt}
\end{align}
where $p(\bfx_{t_{k}}|\bfz^{t_{k-1}})$ is the predictive\footnote{The predictive PDF $p(\bfx_{t_{0}}|\bfz^{t_{-1}})$ is equal to the initial one  $p(\bfx_0)$.} PDF, $p(\bfz_{t_k}|\bfx_{t_k})$ is the measurement PDF obtained from \eqref{eq:meas}, and the normalization coefficient is given by 
	$p(\bfz_{t_k}|\bfz^{t_{k-1}})=\int p(\bfx_{t_k}|\bfz^{t_{k-1}})p(\bfz_{t_k}|\bfx_{t_k})d\bfx_{t_k}$.

\subsection{Time Update}
The solution to the time update, w.r.t. the SDE \eqref{eq:dynam}, is given by the Fokker-Planck equation (FPE) describing the time evolution of the filtering PDF to predictive PDF
\begin{align}
\pderiv{p_t(\bfx|\bfz^{t_k})}{t} &= -\nabla\cdot \left(\bfA \bfx p_t(\bfx|\bfz^{t_k}) \right) \nonumber\\
& + \frac{1}{2}\nabla\cdot\left(\bfQ\nabla p_t(\bfx|\bfz^{t_k}) \right), \label{eq:fokker}
\end{align}
where $t\in(t_k,t_{k+1})$, the operator $\nabla$ denotes the gradient w.r.t. $\bfx$, and $\nabla\cdot$ stands for the divergence. Further, when possible, the conditional PDF is written simply as $p_t(\bfx)$. 

In  \eqref{eq:fokker}, the first right-hand side term is named \textit{hyperbolic} and it describes the \textit{advection} of the PDF tied to the state dynamics. The second term is named \textit{parabolic} and it describes the \textit{diffusion} caused by the state noise. The standard FPE \eqref{eq:fokker} holds for the Gaussian state noise only \cite{DeHoHa:09}.

\section{FPE Efficient Solution Prerequisites}

The FPE is exactly solvable for a narrow set of the SDEs \eqref{eq:dynam}. Within the set, it is possible to find the linear SDEs, for which the solution to the FPE \eqref{eq:fokker} leads to the prediction step of the Kalman-Bucy filter \cite{Ja:70}, or SDEs of special form that are solved by e.g. Beneš, Daum, and Wong and Yau filters \cite{Da:05}. For remaining SDEs, the FPE is not exactly solvable and need to be solved either using model linearisation or numerically on a grid of points \cite{risken1989fpe,Ja:70,NgPfDe:05}. 


Before the numerical solution is discussed, the FPE is rewritten into suitable form allowing simple application of fast frequency-based numerical methods. The changes are twofold, first, a diagonalisation of matrix $\bfQ$ is performed, second, an advection is solved using Lagrangian approach.

\subsection{Diagonalization}
 First step that greatly reduces the complexity of the advanced efficient numerical algorithms is a diagonalisation of the $\bfQ$ in \eqref{eq:dynam} and subsequently in \eqref{eq:fokker}. Instead of solving the FPE for the model with non-diagonal $\bfQ$, an alternative diagonalized FPE can be solved
	\begin{align}
 	{\pderiv{p_{t}(\bar{\bfx})}{t}} &= -{\nabla_{\bar{\bfx}}\cdot \left(\bar{\bfA} \bar{\bfx}\ p_{t}(\bar{\bfx}) \right)}\nonumber\\
 	& + \frac{1}{2}\nabla_{\bar{\bfx}}\cdot\left(\nabla_{\bar{\bfx}}^T p_{t}(\bar{\bfx})\right), \label{eq:fokker2}
\end{align}
where the following substitution was done, $\bfQ = \bfG \bfG^T$, $\bar{\bfx}= \bfG^{-1} \bfx$, $\bar{\bfA} = \bfG^{-1}\bfA\bfG$, and $\nabla_{\bar{\bfx}} = [ \frac{\partial}{\partial \bar{\bfx}(1)} ... \frac{\partial}{\partial \bar{\bfx}(n_x)}]$. Then the estimation algorithm is run in the state space $\bar{\bfx}$. If the first two moments (i.e. the estimate and its uncertainty) for the original FPE solution are needed they can be calculated as $\mean[\bfx] = \bfG\mean[\bar{\bfx}]$, $\cov[\bfx] = \bfG\ \cov[\bar{\bfx}]\ \bfG^{-1}$.  For simplicity it is further assumed that $\bfQ$ is diagonal from the very beginning, which is true for many models such as the model in the numerical illustration section of this paper.

\subsection{Advection solution}
The second step applies a Lagrangian approach to solve the advection part of the FPE $ \nabla p_t(\bfx) \left(\bfA \bfx\right)$ \eqref{eq:fokker}. This part is problematic because the product of $p_t(\bfx)$ and $\bfA \bfx$ depends on the actual state, which prevents efficient solution in frequency domain. This part can be disposed of by using - in a sense - a Lagrangian approach \cite[pp.~7]{Ma:19}.

	Let every point of the state-space $\bfx\in\real^{n_x}$ be moving according to\footnote{Movement of the PDF caused by the advection is compensated by the movement of the observer. Imagine you are one particle of water in the river, in a simplified way, from your point of perspective, the water around you is not moving.} $\dot{\bfx} = \bfA \bfx$. Then, the FPE advection part \eqref{eq:fokker} taking into account the movement reads \cite{LiPe:97}
	\begin{align}
	\nabla\cdot  \left(\bfA \bfx\ p_t(\bfx)  \right) =  \nabla p_t(\bfx)\ \dot{\bfx} +
	\operatorname{trace}(\bfA) p_t(\bfx),
	\label{eq:movFPEadv}
	\end{align} 
	where $\operatorname{trace}(\bfA)$ is the trace of the matrix $\bfA$.

	The FPE to be solved is then
	\begin{align}
	{\pderiv{p_t(\bfx)}{\bfv}} &= - \operatorname{trace}(\bfA)p_t(\bfx)   \nonumber\\
	& +\frac{1}{2}\nabla\cdot\left(\bfQ\left(\nabla^T p_t(\bfx)\right) \right),\label{eq:easyFPE}
	\end{align}
	where $\bfv = \begin{bmatrix}
	\bf{\dot{x}} & 1
	\end{bmatrix} $, and \begin{align}
		{\pderiv{p_t(\bfx)}{\bfv}} = \bfv \begin{bmatrix}
	\nabla p_t(\bfx) \ ,& \ \pderiv{p_t(\bfx)}{t}
	\end{bmatrix}^T.
	\end{align}
	
	Compared to the FPE  \eqref{eq:fokker}, where the PDF is differentiated w.r.t. the time $t$, in \eqref{eq:easyFPE} it is differentiated w.r.t. observer, which is moving in time. As will be seen, the remaining advection term $\operatorname{trace}(\bfA)p_t(\bfx)$ can be easily handled.
	 
	\section{Grid-Based Estimation}
	In this section a grid-based estimation for continuous dynamics is briefly introduced, starting with usually used approximation to the PDFs. After that two grid-based estimation methods are presented; \textit{(i)} standard basic finite difference method in state domain, and \textit{(ii)} is its efficient formulation in frequency domain. 
	
	Methods, are presented in their one-dimensional form for better understanding of the underlying concepts. Multidimensional variants can be found in \cite{MaDuBr:23,DuMaSt:23,MaDuBrCh:23}.
	
	The reason for using grid-based methods over mesh-less methods such as particle methods is that grid based methods offer deterministic results and are believed to be more robust \cite{AnHA:06}.
	
	\subsection{Density Approximation}
	
	The grid-based filters are based on an approximation of a conditional PDF  $p_t(\bfx)$ by a \textit{piece-wise constant} point-mass density (PMD) $\hat{p}_t(\bfx;\bfxi_t)$ defined at the set of the discrete grid points\footnote{MATLAB\textregistered\ / Python\textregistered\ style notation is used throughout the paper for simple comparison with published codes, where  $(j+1)$-th element of vector $\bfx_t$ is denoted as $\bfx_t^{(j)}$ and element of matrix $\bfxi_t$ at $(i+1)$-th row and $(j+1)$-th column is denoted as $\bfxi_t^{(j,i)}$. The indexing is from $0$ as it allows transparent notation for spectral differentiation theory.} $\bfxi_t=[\bfxi^{(:,0)}_t,...,\bfxi^{(:,N-1)}_t], \bfxi^{(:,i)}_t\in\real^{n_x}$, as follows 
\begin{align}
\hat{p}_t(\bfx;\bfxi_k)\triangleq\sum_{i=0}^{N-1}P_t(\bfxi^{(:,i)}_t)S_t\{\bfx;\bfxi^{(:,i)}_t,\bfdelta_t\},\label{eq:PDF_pm}
\end{align}
with
\begin{itemize}[leftmargin=\parindent,align=left,labelwidth=\parindent,labelsep=0pt,noitemsep]
	\item $P_t(\bfxi^{(:,i)}_t)=c_t\tilde{P}_t(\bfxi^{(:,i)}_t)$, where $\tilde{P}_t(\bfxi^{(:,i)}_t)=p_t(\bfxi^{(:,i)}_t)$ is the value of the conditional PDF $p_t(\bfx)$ evaluated at the $i$-th grid point $\bfxi^{(:,i)}_t$ (also called PMD point weight), $c_t=\delta_t\sum_{i=1}^{N}\tilde{P}_t(\bfxi^{(:,i)}_t)$ is a normalisation constant, and $\delta_t$ is the volume of the $i$-th point neighbourhood,
	\item $\bfdelta_t$ defines a (hyper-) rectangular neighbourhood\footnote{Assumed to be same for each grid point, i.e. the grid is equidistant.} of a grid point, where the PDF $p_t(\bfx)$ is assumed to be constant and has value $P_t(\bfxi^{(:,i)}_t)$,
	\item $S_t\{\bfx;\bfxi^{(:,i)}_t,\bfdelta_t\}$ is the \textit{selection} function defined as
	\begin{align}
	S_t\{\!\bfx;\bfxi^{(:,i)}_t\!,\!\bfdelta_t\!\}\!=\!\begin{cases}
	\!1,\mathrm{if}\ |\bfx^{(j)}\!-\!\bfxi^{(j,i)}_t|\!\leq\!\tfrac{\bfdelta_t^{(j)}}{2}\forall j,\\
	\!0, \mathrm{otherwise}.
	\end{cases}\!\!\!\!\label{eq:sf3}
	\end{align}
	\item $N = (N_\text{pa})^{n_\bfx}$ is the total number of grid points and $N_\text{pa}$ means the number of discretisation points per axis (pa) for grid with boundaries aligned with the state-space axes.
	\end{itemize}
	
	\subsection{Finite Difference Method}
	
    The finite difference method (FDM) can be considered as the baseline approach typically used in the continuous grid-based state estimation, where the continuous conditional PDF support is approximated by the grid points in which the differences are computed to solve the FPE.  
  

    The number of possible FDM numerical schemes is vast, and their overview can be e.g. found in \cite{LeVe:07}. The baseline approach in the state estimation is an explicit scheme with central difference for diffusion (upwind for advection). The FPE \eqref{eq:easyFPE} is solved from $t_k$ to $t_{k+1}$ using $l$ numerical steps with length $\Delta t$, i.e., $l = \frac{t_{k+1} - t_k}{\Delta t} $.
 
        For the FPE \eqref{eq:easyFPE}, and the enforced grid movement 
        \begin{align}
        	\dot{\xi}_{n}^{(j)} = A {\xi}_{n}^{(j)}, \forall j, \label{eq:gridMov}
        \end{align} 
        an explicit 1D scheme can be derived, using shorthand notation $P^{(j)}_{n} = P_{t_k+n\Delta t}(\xi^{(j)}_{t_k+n\Delta t})$, as
   \begin{align}
   	P^{(j)}_{n+1} = P^{(j)}_{n} - A P^{(j)}_{n}  + \nonumber\\\frac{\Delta t}{2\delta_n^2}Q \left(  P^{j+1}_n - 2P^{(j)}_n + P^{j-1}_n\right), n=0,1,\ldots,l, \label{eq:FDM}
   \end{align}
   where $\delta_n=e^{A\Delta t}\delta_{n-1}$. 
    For multi-dimensional case, a derivative in each direction is approximated by the difference, see e.g. \cite{LeVe:07}.
	
	\subsection{Efficient Formulation}
	
		In standard formulation, to calculate the advection part of the FPE \eqref{eq:fokker} a product of dynamics $A x$ and PDF $p_t(x)$ has to be done. In practice this equals to Hadamard product of two vectors. Unfortunately, Hadamard product in the time domain becomes convolution in the frequency domain. Therefore, for each numerical time step, the PDF derivative has to be converted back to the time domain, the product calculated and the result converted back to the frequency domain, leading to computational overhead. However, the efficient FPE formulation  \eqref{eq:easyFPE} circumvents this issue as it solves the advection term by a grid movement.
	
	The finite difference solution \eqref{eq:FDM} can be rewritten to a matrix form, forming a tridiagonal matrix
	\begin{align}
	\bfF_{\text{diff}}(t) = \begin{bmatrix}
	b_t  & a_t &  &   \\
	a_t & b_t  & a_t & & \  \\
	& a_t & b_t & a_t &    & \\
	&  & \ddots & \ddots &  \ddots   & \\
	\end{bmatrix}, \label{eq:fdiff}
	\end{align}
	where $a_t = \frac{Q\ \Delta t}{2\delta_t^2}$, and $b_t = 1 - \frac{Q \Delta t}{\delta_t^2} - \Delta tA$. Note that the coefficients $a_t, b_t$ are row-independent due to the FPE with enforced grid movement \eqref{eq:easyFPE}.

	The resulting prediction, i.e., a numerical solution to the FPE \eqref{eq:easyFPE}, from the time instant $k$ to $k+1$, becomes
	\begin{align}
	P_{{t_{k+1}}}^{(:)} =\underbrace{\bfF_{\text{diff}}(t_k+l\Delta t) \cdots \bfF_{\text{diff}}(t_k+\Delta t)\bfF_{\text{diff}}(t_k)}_{\bfT_k}\ P_{{t_k}}^{(:)}, \label{eq:FKEnumSol}
	\end{align}
	where the shorthand notation $P_{{t_{k}}}^{(:)}=\left[P_{{t_{k}}}^{(0)}, P_{{t_{k}}}^{(1)}, \ldots,\right.$ $\left. P_{{t_{k}}}^{(N-1)}\right]^T$ is used.
	
	It can be shown that thanks to the grid movement \eqref{eq:gridMov}, the eigenvector matrix $\bfR$ of time dependent $\bfF_{\text{diff}}(t)$ are constant, while its eigenvalues $\lambda_t$ are time-dependent. Thus $\bfT_{t_k}$ can be, efficiently, calculated using the eigenvalue and eigenvector form of $\bfF_{\text{diff}}(t)$ as
		\begin{align}
	\bfT_{t_k} &= \bfR \nonumber\\ &\underbrace{\left(\bfLambda(t_k+(l-1)\Delta t)\odot \cdots \odot\bfLambda(t_k+\Delta t) \odot \bfLambda(t_k) \right)}_{\bfLambda_{t_k}}\bfR^{-1},
	\end{align}	
	where $\bfLambda_k$ is a matrix with eigenvalues $\lambda_t^{(j)}$ on diagonal. Because the matrix $\bfF_{\text{diff}}$ \eqref{eq:fdiff} is Toeplitz, its eigenvalues are \cite{Sa:06}
	\begin{align}
	\lambda^{(j-1)}_t = b_t + 2a_t \cos\left(\frac{j\pi}{N+1} \right), j=1,\ldots,N,
	\end{align}
	and eigenvectors
	\begin{align}
	\bfr^{(j-1)} = \begin{bmatrix}
	\sin(\frac{1j\pi}{N+1})\\
	\vdots \\
	\sin(\frac{Nj\pi}{N+1})\\
	\end{bmatrix},\ j=1,...,N.
	\end{align}
	For the eigenvector matrix, it also holds that 
	\begin{align}
	\bfR^{-1} = \frac{2}{N+1} \bfR.
	\end{align}	
	Therefore, calculation of an arbitrary element $\bfT^{(j-1,i-1)}_{t_k}$ reads
	\begin{align}
	\bfT^{(j-1,i-1)}_{t_k} = \frac{2}{N+1} \sum_{\ell=1}^N \bfLambda_{t_k}^{(\ell,\ell)} \sin\left( \frac{i\ell\pi}{N+1} \right) \sin\left( \frac{j\ell\pi}{N+1} \right),
	\end{align}
	where $i=1,\ldots,N$ and $j=1,\ldots,N$. Note that calculation of the element $\bfT^{(j-1,i-1)}_{t_k} $ is independent of any other element compared to the standard FDM-based \eqref{eq:FDM} solution. Thanks to this special form, the \textit{fast sine transform} $\mathcal{S}$ can be used \cite{St:07} to calculate prediction as
	\begin{align}
			 P_{{t_{k+1}}}^{(:)} = \mathcal{S} \left( {\diag(\bfLambda_{t_k})} \odot \mathcal{S}({ P}_{{t_k}}^{(:)}) \right), \label{eq:FST}
	\end{align}
where $\odot$ is Hadamard product. For further information and multidimensional derivation please refer to \cite{MaDuBr:23,DuMaSt:23,MaDuBrCh:23}.

	 \subsection{Computational Complexity and Convergence Rate}
The computational complexity of predictive PDF calculation in grid-based method was reduced from $\mathcal{O}(N^2)$ in \eqref{eq:FDM} to $\mathcal{O}(N \log N)$ in \eqref{eq:FST}. However, the convergence rate is still $\mathcal{O}(\frac{1}{N^2})$. 

Thus, the aim of this paper is to keep the reduced computational complexity and also enhance the convergence rate in space, i.e., to improve FPE solution with reduced number of grid points $N$.
	
	\section{Spectral differentiation}
	Spectral differentiation (SD) is a method of calculating derivative of a function in a frequency domain, such as derivatives present in the FPE \eqref{eq:easyFPE}. The main advantage of the spectral methods is very fast convergence rate $\mathcal{O}(c^{N}) (O< c < 1)$ for analytic functions \cite[pp.~41]{Th:00}. 
	
	Basic idea of the SD in this case is, instead of calculating the conditional PMD differences in the state-space domain, to interpolate the Fourier transformation of the PMD and calculate its derivative analytically. 
		
	Let a one dimensional PDF $p_t(x)$ be sampled on a grid with $N$ grid points $\xi_t^{(j)}$, where $N$ is assumed to be even, and let $L_t=N\delta_t$ be a grid size. Then, the  PMD can be expressed in a frequency domain using discrete Fourier transform
	\begin{align}
		\mathcal{P}_t^{(s)} = \frac{1}{N} \sum_{j=0}^{N-1}  P_t^{(j)} \exp^{-\frac{2\pi i}{N}js}, \label{eq:fourtrsf}
	\end{align}  
	where $s=0,\ldots,N-1$ and $i$ is imaginary unit.
	
	To compute derivatives of \eqref{eq:fourtrsf} analytically, a continuous interpolation in the frequency domain has to be defined. A unique "minimal-oscillation" trigonometric interpolation of order $N$ is \cite{Jo:11}
	\begin{align}
		p_t(x) = & \mathcal{P}_t^{(0)} + \sum_{0<s<N/2}  \left( \mathcal{P}_t^{(s)} e^{\frac{2\pi i }{L_t}sx} + \mathcal{P}_t^{(N-s)} e^{-\frac{2\pi i }{L_t}sx} \right) + \nonumber\\ 
		& \mathcal{P}_t^{(N/2)} \cos\left( \frac{\pi}{L_t}Nx \right). \label{eq:interFreq}
	\end{align}
	Interpolated PMD in frequency domain \eqref{eq:interFreq} can easily be differentiated w.r.t. state, $\pderiv{p_t^{(j)}}{x} = p_t'^{(j)}$ and evaluated at each grid point
	\begin{align}
		p_t'^{(j)} &= \sum_{0<s<N/2} \underbrace{\frac{2\pi i }{L_t}s}_{c'_t} \left( \mathcal{P}_t^{(s)} e^{\frac{2\pi i }{N}js} + \mathcal{P}_t^{(N-s)} e^{-\frac{2\pi i }{N}js} \right)\nonumber\\
		& = \sum_{s=0}^{N-1} \mathcal{P}_t'^{(s)} \exp^{\frac{2\pi i}{N}js}.
	\end{align}
	The second derivative $\pderivv{p_t^{(j)}}{x} = p_t''^{(j)}$, needed for the diffusion solution, is then given as
		\begin{align}
		p_t''^{(j)}\!\!\!&=\!- \!\!\!\!\!\! \sum_{0<s<N/2} \underbrace{\left(\frac{2\pi i }{L_t}s \right)^2}_{c''_t}\left( \mathcal{P}^{(s)}_t e^{\frac{2\pi i }{N}js} + \mathcal{P}_t^{(N-s)} e^{-\frac{2\pi i }{N}js} \right)  \nonumber\\
		&- \left( \frac{\pi}{L_t}N \right)^2 \mathcal{P}_t^{(N/2)} (-1)^j		\nonumber\\ 
		&=\sum_{s=0}^{N-1} \mathcal{P}_t''^{(s)} \exp^{\frac{2\pi i}{N}js}, \label{eq:secDer}
	\end{align}
	
	Applying the derivatives in multiple dimensions is done by applying successive one dimensional fast Fourier transforms and differentiations. This will be shown later in algorithmic form.
	
	\subsection{Illustration of Differentiation}
\begin{figure}[]
		\centering
		\includegraphics[width=1\linewidth]{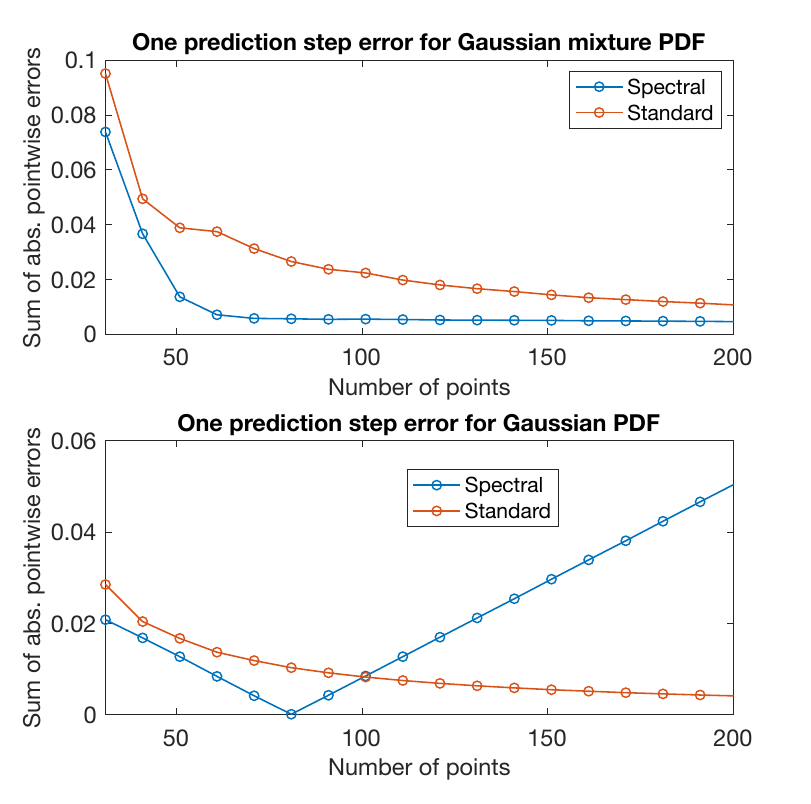}\vspace*{-4mm}
		\caption{Time update accuracy: convergence for Gaussian PDF and Gaussian mixture PDF.}
		\label{fig:convergence1}\vspace*{-4mm}
	\end{figure}

	 In Figure \ref{fig:convergence1}, an error of the time update step using the FPE \eqref{eq:easyFPE} for model \eqref{eq:dynam} without advection term (for the sake of simplicity) can be seen for the filtering PDF $ p(\bfx_{t_k}|\bfz^{t_k})$ in the form 
	 \begin{itemize}
	 	\item the Gaussian mixture with three components,
	 	\item the Gaussian PDF.
	 \end{itemize}
	 The fast convergence of the spectral method compared to standard FDM is evident. However, in the case that underlying PDF's are Gaussian a strange behaviour is observed, where the accuracy starts deteriorate with more points. Similar behaviour can be seen for example in \cite{DuDu:22} without any commentary. Note that, this is not caused by violating the Courant–Friedrichs–Lewy criterion because similar behaviour can be seen in the error of derivative approximation itself.
	
	Our hypothesis is that this behaviour is caused by the trigonometric interpolation as it assumes the function is periodic, please see next section for details.Therefore, in future research we would like to try using the periodic $sinc$ function for interpolation \cite[pp.~20]{Th:00}. 
	
	The trigonometric interpolation was used nonetheless as it is simple and shows good performance, as the considered model with advection \eqref{eq:dynam}, \eqref{eq:meas} is nonlinear and the underlying conditional PDFs are, therefore, inherently non-Gaussian. 
	
\begin{figure}[]
		\centering
		\includegraphics[width=1\linewidth]{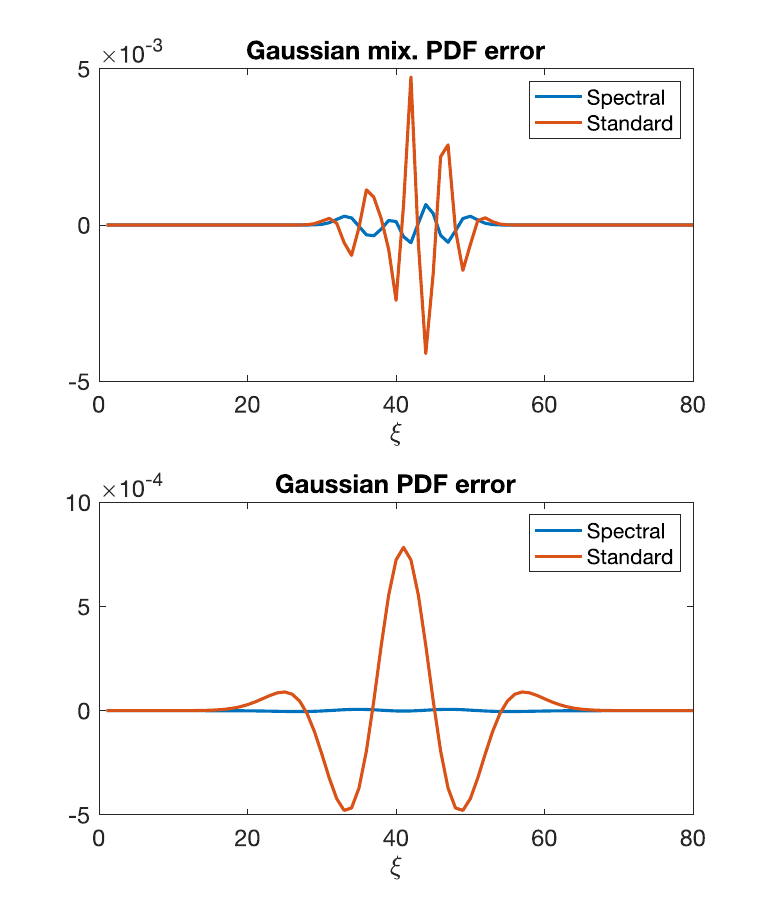}\vspace*{-4mm}
		\caption{Predictive PMD error for $N=80$.}
		\label{fig:convergence2}\vspace*{-4mm}
	\end{figure}

\begin{figure}[]
		\centering
		\includegraphics[width=1\linewidth]{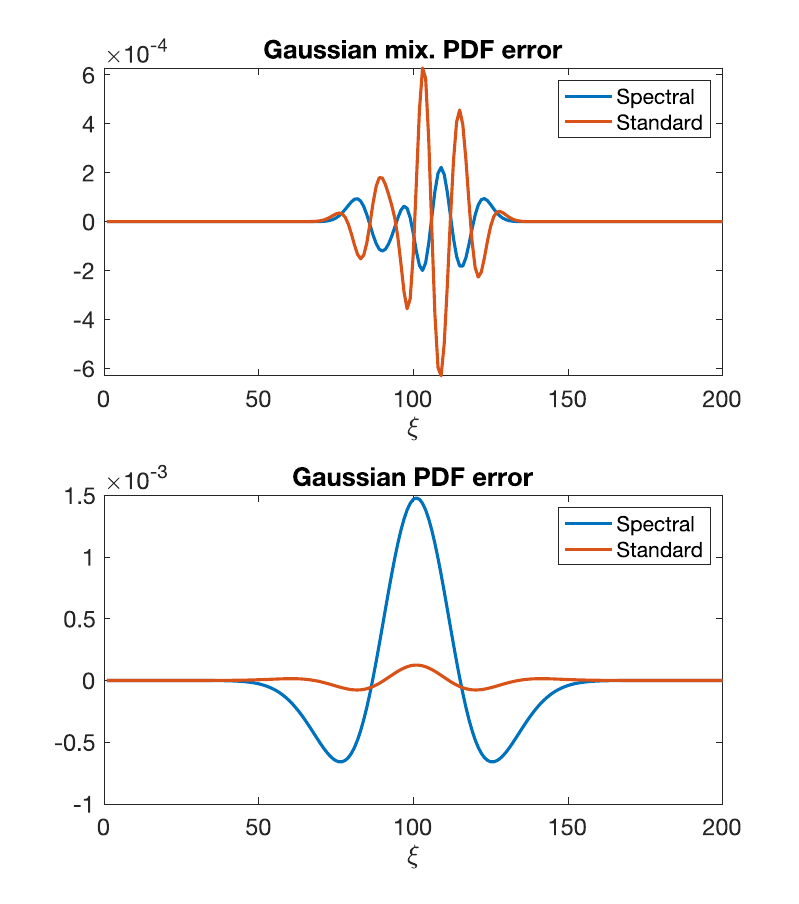}\vspace*{-4mm}
		\caption{Predictive PMD error for $N=200$.}
		\label{fig:convergence}\vspace*{-4mm}
	\end{figure}

	Illustration of the errors of the calculated predictive PDF $p_{t_{k+1}}(\bfx|\bfz^{t_k})$ for $N=80$ and $N=200$ is shown in Figures \ref{fig:convergence2}  and \ref{fig:convergence}, respectively,  for both filtering densities. It can be seen that for  $N=80$, the SD-based FPE solution provides significantly better prediction for both filtering PDFs. However, for  $N=200$ the accuracy of the prediction using the SD-based FPE solution is worse for the Gaussian filtering PDF as discussed above.

	\subsection{Notes}
	    Considered SD method is intended for differentiation of the periodic functions, however, PDFs are not mathematically periodic.  Fortunately, when the grid is well designed, the underlying PDF should always be near zero at the boundaries of the grid in such case it can be viewed as periodic in practice \cite[pp. 24]{Th:00}.

		As an alternative for non-periodic functions that are not near zero at the boundaries of the considered domain, a Chebyshev interpolation and differentiation can be used \cite[pp.~41]{Th:00}. However, the grid points have to be Chebyshev nodes, such a grid has complex structure and is not suitable for state estimation.
	
		There is a number of other interpolations and approaches that can be use in spectral differentiation, but for the sake of this paper, which has aim to prove the superiority of spectral methods for state estimation in general, the described approach was chosen.

	\section{Spectral Based Estimation}
	
		\begin{figure}
		\includegraphics[width=0.5\textwidth]{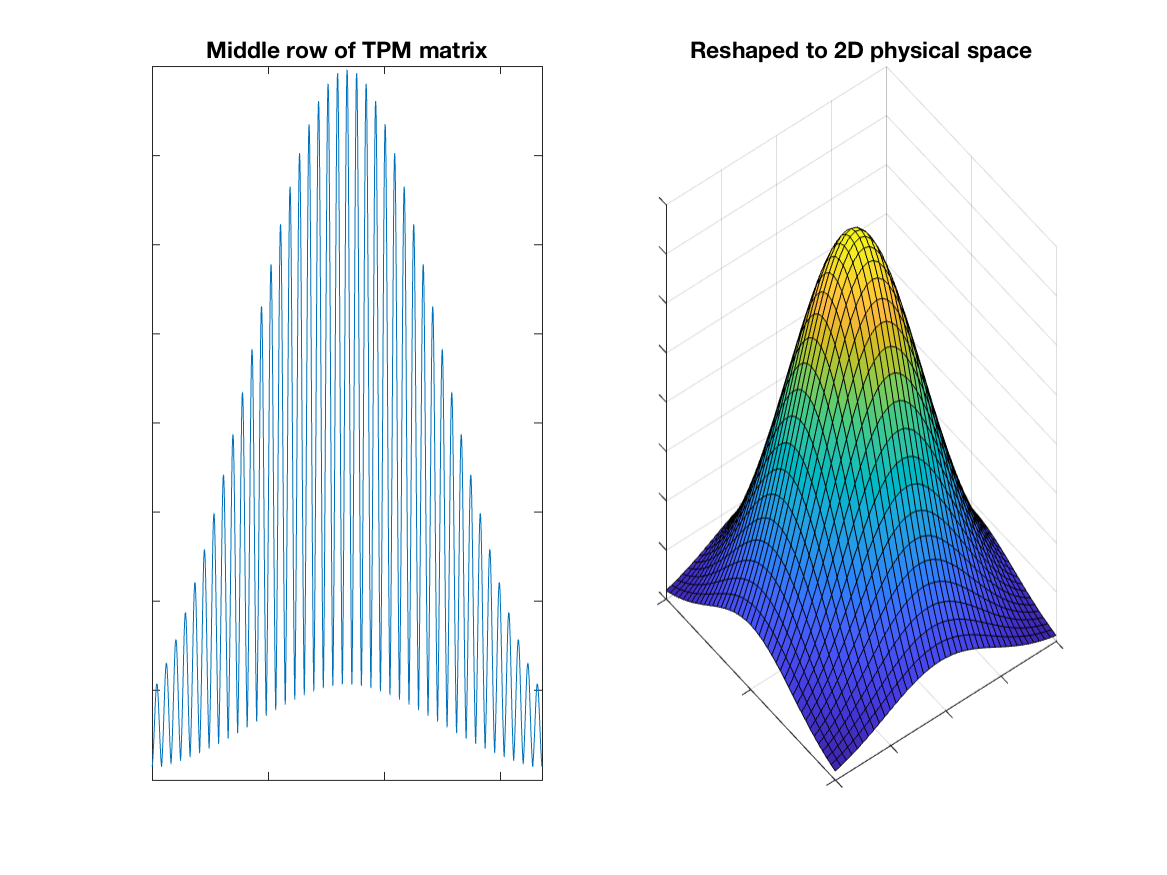}
		\caption{Computational to physical space re-shaping for 2D space and Gaussian noise.}
		\label{fig:reshape}
	\end{figure}
	
After the advection was solved by Lagrangian approach and grid movement \eqref{eq:easyFPE}, what remains is to treat the diffusion term
		\begin{align}
		 \frac{1}{2}\nabla\cdot\left(\bfQ\nabla p_t(\bfx(t) \right),
	\end{align} 
 where second derivative of $p_t(\bfx)$ is to be calculated. Because $\bfQ$ is constant and diagonal, the product with $p_t(\bfx)$ can be easily done in a frequency domain.
 
 For simplicity the equations are derived for $n_x=1$ and the extension for arbitrary $n_x$ is given in the form of the algorithm.
 
 \subsection{Numerical Time Stepping with Spectral Methods}
 
 The coefficients to calculate the second derivate, from \eqref{eq:secDer}, are then
 \begin{align}
 	c''_t = -\left(\frac{2\pi}{L_t} \text{fftshift}\left(-\frac{N_{\text{pa}}}{2}:1:(\frac{N_{\text{pa}}}{2}-1) \right)\right)^2,
 \end{align}
 where MATLAB function \verb|fftshift| shifts the zero frequency component to the middle, and colon follows MATLAB notation.
	The numerical time stepping including the $\operatorname{trace}$ can be, for simplicity (as our main aim here is to show the spectral method), done using Euler discretisation time step as \cite{Ue09}
	\begin{multline}
		\frac{1}{\Delta_t} \left( \mathcal{P}_{n+1} -  \mathcal{P}_{n} \right) = - \frac{q}{2} c''_n \odot \mathcal{P}_{n+1} + 	\operatorname{trace}(\bfA) \mathcal{P}_{n} \nonumber\\
		\Leftrightarrow  \mathcal{P}_{n+1} = \left({\mathcal{P}_{n} (1+\operatorname{trace}(\bfA))}\right) \oslash \left(1 - \frac{q}{2}\Delta_t c''_n\right)\nonumber\\ = {\mathcal{P}_{n}(1+\Delta_t\operatorname{trace}(\bfA))}\oslash {\psi_{n}},
	\end{multline}
	where $\oslash$ is a Hadamard division.
	
\subsection{Spectral Methods State Estimation Algorithm}	
The point-mass filter prediction step based on the introduced SD method and Euler stepping is summarised in the Algorithm below for an arbitrary state dimension $n_x$.

	\noindent\rule{8.5cm}{0.5pt}
\newline
\textbf{Algorithm: Efficient spectral point-mass filter weight update}
\vspace*{-1mm}
\begin{enumerate}
	\setcounter{enumi}{0}
	\item \textit{Calculate fast Fourier transform of the filtering PMD weights $P_{t_k|t_k}$ dimension by dimension}: $ \widetilde{\mathcal{P}}_{t_k|t_k} = ... \mathcal{F}_{x_2}(\mathcal{F}_{x_1}(\widetilde{P}_{t_k|t_k}))$
	\item \textit{Calculate coefficients for gradient}:\\ $c''^{(1,:)}_{t_k} = -\left(\frac{2\pi}{L_{t_k}^{(1)}} \textit{fftshift}\left(-\frac{N_{\text{pa}}}{2}:1:(\frac{N_{\text{pa}}}{2}-1) \right) \right)^2$\\
	  $c''^{(2,:)}_{t_k} = -\left(\frac{2\pi}{L_{t_k}^{(2)}} \textit{fftshift}\left(-\frac{N_{\text{pa}}}{2}:1:(\frac{N_{\text{pa}}}{2}-1) \right)\right)^2\\
	  \hspace*{1cm}\vdots$
	\item \textit{Calculate (generally) tensor $\Psi_{t_k}$ as a product $\psi_{t_k} \odot \psi_{t_k + \Delta_t} \odot \psi_{t_k + 2\Delta_t} \dots \odot \psi_{t_{k+1}}$ where}: 
	\begin{align}
		\psi_{t_k}^{(i,j,...)} = \tfrac{1}{1 - 0.5c''^{(1,i)}_{t_k}Q^{(1,1)}\Delta_t - 0.5c''^{(2,j)}_{t_k}Q^{(2,2)}\Delta_t-\cdots}\nonumber
	\end{align}
	\item \textit{Calculate the updated weights}: \\$\mathcal{P}_{t_{k+1}|t_k} = \Psi_{t_k} \odot \mathcal{P}_{t_k|t_k}(1 + \Delta_t \operatorname{trace}(\bfA))^l$
	\item \textit{Calculate inverse fast Fourier transform of $\mathcal{P}_{t_{k+1}|t_k}$}
	\item \textit{Normalize} 
\end{enumerate}
\vspace*{-3mm}
\rule{8.5cm}{0.5pt}
\newline
\vspace*{-3mm}

Note that, in the algorithm, the terms with $\sim$ overhead are reshaped to physical space as shown in Figure \ref{fig:reshape} as discussed in \cite{MaDuBrCh:23, DuMaSt:23, MaDuBr:23}, and  $L_{t_k}^{(m)}$ is the size of the grid for $m$-th dimension.
	
	\section{Numerical Illustration}
	A model of dynamic coordinated turn with known turn rate $\alpha$ was used \cite{LiJi:03}. A four-dimensional state $\bfx_k = [p_x \ v_x \ p_y  \ v_y]$, which describes the horizontal position ($p_x, p_y$) $[m]$ and velocity ($v_x, v_y$) $[m/s]$ of the vehicle. 

A continuous dynamics model reads
	\begin{align}
				\bfA &=  \begin{bmatrix}
			0 & 1 & 0  & 0 \\
			0 & 0 & 0 & -\alpha \\
			0 & 0 & 0  & 1 \\
			0 & \alpha & 0  & 0 \\
		\end{bmatrix},\\
		\bfQ &=  \begin{bmatrix}
			0 & 0 \\
			1 & 0  \\
			0 & 0 \\
			0 & 1 \\
		\end{bmatrix},
	\end{align}
	where $\alpha = 30^{\circ}$ is the known turn rate.
	
	An analytically derived discrete dynamics model, used for comparison is, 
		\begin{align}
			\bfx_{k+1}&=\bfF \bfx_{k} + \bfw_{k}\\
		\bfF &=  \begin{bmatrix}
			1 & \frac{\sin(\alpha T_s)}{\alpha} & 0  & \frac{\cos(\alpha T_s) - 1}{\alpha} \\
			0 & \cos(\alpha T_s) & 0 &-\sin(\alpha T_s) \\
			0 & \frac{1-\cos(\alpha T_s)}{\alpha} & 1 & \frac{\sin(\alpha T_s)}{\alpha}\\
			0 & \sin(\alpha T_s) & 0 & \cos(\alpha T_s)
		\end{bmatrix}\\
		p(\bfx_0) &\sim \mathcal{N}\left\{\bfx_0;\begin{bmatrix}
			36569 \\
			50 \\
			55581\\
			50
		\end{bmatrix}, \begin{bmatrix}
			90 & 0 & 0 & 0 \\
			0 & 160 & 0 & 0 \\
			0 & 0 & 5 & 0 \\
			0 & 0 & 0 & 5 
		\end{bmatrix} \right\}, \\
		p(\bfw_k) &\sim \mathcal{N}\{ \mathbf{w}_k ;\mathbf{0}, \bfQ_d \},
		\end{align} \begin{align} \\\bfQ_d
 &=
\left[\begin{matrix}
 \frac{2(\alpha T_s - \sin(\alpha T_s))}{\alpha^3} & \frac{1-\cos(\alpha T_s)}{\alpha^2}\\
 \frac{1- \cos(\alpha T_s)}{\alpha^2} & T\\
  0 & \frac{-\alpha T_s - \sin(\alpha T_s)}{\alpha^2}\\
  \frac{\alpha T_s - \sin(\alpha T_s)}{\alpha^2} & 0
  \end{matrix}\right.\nonumber\\
&\qquad\qquad
\left.\begin{matrix}
  0 & \frac{\alpha T_s - \sin(\alpha T_s)}{\alpha^2}\\
  -\frac{\alpha T_s - \sin(\alpha T_s)}{\alpha^2} & 0\\
  \frac{2(\alpha T_s - \sin(\alpha T_s))}{\alpha^3} & \frac{1- \cos(\alpha T_s)}{\alpha^2}\\
  \frac{1-\cos(\alpha T_s)}{\alpha^2} & T_s
\end{matrix}\right], \label{eq:asx2}
	\end{align}
	where, $T_s = 1$ is the time step.	
	
	Note the complexity of both models, at first sight it might be beneficial to use the continuous dynamics model due to its simple structure, $Q$ is diagonal, $\operatorname{trace}(\bfA) = 0$, noise is in two state variables only...

		 The measurement equation is the same for both models. The measurement function $h$ is a discrete terrain map\footnote{The map is from Shuttle Radar Topography Mission (SRTM) an international project spearheaded by the U.S. National Geospatial-Intelligence Agency (NGA) and the U.S. National Aeronautics and Space Administration (NASA), see https://www2.jpl.nasa.gov/srtm/index.html.} represented by a table function that assigns vertical position (i.e. altitude) to each combination of latitude and longitude it covers. The measurement $z_k$ is a terrain altitude below the vehicle which can be based on a barometric altimeter. 
	 
	 The noise $v_k$ is distributed according to Gaussian mixture PDF with two components (this is simulating terrain with unmapped bridge or tunnel)\cite{NoGu:09}
	\begin{align}
	p(v_k)=\sum_{g=1}^{2}\frac{1}{2} \mathcal{N}\{v_k;\hat{v}_g,P_g\},\label{eq:ex1_PDF}
	\end{align}
 where the particular means and covariance matrices are given as follows
$\hat{v}_1 = 0, \hat{v}_2 = 20,$ $P_1 = P_2 = 1.$

The results can be seen in Table \ref{tab:res} for 50 Monte-Carlo simulations for three filters
\begin{itemize}
	\item Efficient discrete point-mass filter with $N_\mathrm{pa}=34$  \cite{MaDuBrCh:23, DuMaSt:23, MaDuBr:23},
	\item Discrete bootstrap particle filter with $10^6$ particles \cite{DoFrGo:01},
	\item Designed SD-based continuous point-mass filter with $N_\mathrm{pa}=34$ (Algorithm in Section V.B).
\end{itemize}
The performance of the filters is compared using
\begin{itemize}
	\item RMSE$^{(j)}$ = \begin{align}
		\sqrt{\frac{1}{M(T+1)}\sum_{m=1}^M\sum_{k=0}^{T}((\bfx^{(j)}_{t_k})^{[m]}-(\hbfx^{(j)}_{t_k|t_k})^{[m]})^2},
	\end{align}
	\item ASTD$^{(j)}$ = \begin{align}
		\sqrt{\frac{1}{M(T+1)}\sum_{m=1}^M\sum_{k=0}^{T}{(\bfP^{(j,j)}_{t_k|t_k})^{[m]}}},
	\end{align}
\end{itemize}
	using $M=50$ Monte-Carlo (MC) simulations, where $(\bfx^{(j)}_{t_k})^{[m]}$ is $(j+1)$-th element for the true state at time $k$ and $m$-th MC simulation, $(\hbfx^{(j)}_{t_k|t_k})^{[m]}=\mean[\bfx_{t_k}^{[i]}|\bfz^{t_k}]$ its filtering estimate, and $(\bfP^{(j,j)}_{t_k|t_k})^{[m]}=\var[(\bfx_{t_k}^{(j)})^{[m]}|\bfz^{t_k}]$ the corresponding filtering covariance diagonal element.
	
The table indicates, that the proposed filter provides most accurate and consistent results with the lowest computational complexity.  The higher computational complexity of the discrete efficient filter is due to a need for two fast Fourier transform (FFT) transforms and one inverse FFT (IFFT) as opposed to the efficient spectral continuous filter which needs just one FFT and one IFFT each step. Also, in the discrete case where the FFT is used to perform efficient convolution using the convolution theorem, the FFT is performed on larger tensors (in general), because of the need for zero padding \cite[pp.~27]{IfAb:12}.

Based on the results,  it may be beneficial to use continuous filter as it can be simple, efficient, and accurate.

\begin{table*}
\center
\begin{tabular}{llllllllll}
& RMSE$^{(1)}$ & RMSE$^{(2)}$ & RMSE$^{(3)}$ & RMSE$^{(4)}$ & ASTD$^{(1)}$ & ASTD$^{(2)}$ & ASTD$^{(3)}$ & ASTD$^{(4)}$ & TIME \\ 
\hline 
Efficient discrete & 14.6867 & 13.2812 & 10.071 & 6.9791 & 18.2505 & 14.989 & 12.606 & 8.6158 & 0.82499 \\  
PF bootstrap & 19.0446 & 17.9669 & 11.9933 & 9.2139 & 29.7935 & 29.1996 & 16.862 & 14.1169 & 0.46089 \\
\rowcolor{lightgray}Efficient spectral & 14.4769 & 13.0677 & 9.8878 & 6.8588 & 18.8222 & 15.0611 & 13.1594 & 8.807 & 0.41649 \\ 
\hline 
\end{tabular}
\caption{Results for state estimation in tracking scenario.}
\label{tab:res}
\end{table*}

\section{Concluding Remarks}
The paper dealt with state estimation of continuous in time model with discrete measurement. In particular, the spectral differentiation has been introduced for the efficient solution to the Fokker-Planck equation describing the time evolution of the predictive PDF. This method preserves the computational complexity of efficient formulation of point mass filter $\mathcal{O}(N \log N)$ while achieving superior space convergence rate of $\mathcal{O}(c^{N}) (O< c < 1)$.

 It was shown that the continuous spectral based estimation is less computationally complex and can be defined and implemented in a straightforward way similarly to the discrete grid and particle based state estimation. Despite simple and efficient implementation, the spectral approach leads  to better estimation performance. Moreover, an inherent advantage of the continuous state-space model is its simplicity compared to the discrete counterpart especially in tracking and navigation areas.


In the future research we will focus on more accurate time stepping for spectral methods (i.e., an alternative to Euler method) and on alternative spectral approaches.

\section{Acknowledgment}
J. Dun\'ik, and J. Matou\v{s}ek's work was co-funded by the European Union under the project ROBOPROX - Robotics and advanced industrial production  (reg. no. CZ.02.01.01\slash00\slash22\_008\slash0004590).

	\bibliographystyle{IEEEtran}

\end{document}